# High-Performance Photonic Transformers for DC Voltage Conversion

Bo Zhao[1,3], Sid Assawaworrarit[1,3], Parthiban Santhanam[1], Meir Orenstein[1,2], and Shanhui Fan[1,*]

[1]*Department of Electrical Engineering, Stanford University, Stanford, California, USA*
[2]*Department of Electrical Engineering, Technion Israel Institute of Technology, Haifa, Israel*
[3]*These authors contributed equally.*
*\* To whom correspondence should be addressed. Email: shanhui@stanford.edu.*

Direct current (DC) converters play an essential role in electronic circuits, matching the voltage level of the power source to various voltage levels required by different subcircuits. Conventional high-efficiency DC voltage converters, especially step-up type, rely on switch-mode operation, where energy is periodically stored within and released from inductors and/or capacitors connected in a variety of circuit topologies. However, since these energy storage components, especially inductors, are difficult to scale down, miniaturization of switching converters for on-chip or in-package electronics faces fundamental challenges. Furthermore, the resulting switching currents produce electromagnetic noise, which can cause interference problems in nearby circuits, and even acoustic noise and mechanical vibrations that deteriorate the environment. In order to overcome the limitations of switch-mode converters, photonic transformers, where voltage conversion is achieved through the use of light emission and detection processes, have been demonstrated. However, the demonstrated efficiency is significantly below that of the switch-mode converter. Here we perform a theoretical analysis based on detailed balance, which shows that with a monolithically integrated design that enables efficient photon transport, the photonic transformer can operate with a near-unity conversion efficiency and high voltage conversion ratio. We validate the theoretical analysis with an experiment on a transformer constructed with off-the-shelf discrete components. Our experiment showcases near noiseless operation, **as well as a voltage conversion ratio that is significantly higher than obtained in previous photonic transformer works. Our finding points to a high-performance optical solution to miniaturizing DC power converters for electronics and improving the electromagnetic compatibility and quality of electrical power.**

Voltage conversion plays a critical role in electrical and electronic systems, bridging the gap between the voltage requirements of the power source/generation, distribution, and individual loads on the circuit. As a well-known example that underpins electrical power distribution networks, a transformer (Fig. 1a) converts one alternating-current (AC) voltage to another using the principle of magnetic induction. Such a transformer, however, cannot be applied directly for direct current (DC) voltage conversion.

Since the majority of electronic devices rely on DC power, direct-current (DC) voltage converters are of essential importance in electronics. The standard approach to DC voltage conversion relies on switch-mode converters (Fig. 1b). The core building blocks of these converters are intermediate energy storage elements (inductors and capacitors), and switches (e.g., transistors and diodes) that are temporally modulated to charge and discharge the energy storage elements[1]. Due to the difficulty of scaling down efficient inductors[2], switching converters usually take up substantial real estate on-chip[3] or on the circuit board[4] and therefore represent a major obstacle in miniaturizing electronic devices[5]. Moreover, the switching action inevitably produces fluctuating internal voltages and currents, resulting in significant electromagnetic[6,7] or even acoustic noise[8].

Recently, photon-mediated voltage conversion techniques have been proposed as an alternative solution for DC voltage conversion[9,10] in order to overcome some of the limitations of standard switch-mode converters. In this paper, we refer to such a photon-based voltage converter as a photonic transformer. In a photonic transformer, a laser or a light-emitting diode (LED) is used to convert electrical energy to light energy, and either multiple photovoltaic (PV) cells laterally connected in series or a tandem multi-junction PV cell that are used to convert photon flux back to electrical energy. In doing so, a small input voltage that drives a laser or an LED can be boosted to a larger output voltage from the PV cells. Photonic transformers have been shown to produce no switching noise and are immune to the environment EMI and have been utilized to build DC voltage converters[10] and gate drivers[11]. However, the experimentally demonstrated efficiency, as well as voltage conversion ratio, are significantly below that of the standard switch-mode voltage converter[10]. Therefore, it is important to establish a fundamental understanding of the performance potential of photonic transformers, and the practical pathways to reach such potential.

In this paper, we provide a theoretical analysis of the fundamental performance potential for photonic transformers. We, in particular, focus on a photonic transformer that utilizes LEDs. From a fundamental thermodynamic point of view, the ultimate efficiency of an LED is higher than that of a laser. The electroluminescent efficiency of an LED, theoretically, can in fact exceed 100%[12,13], since an LED can operate as a heat engine that generates part of its light from the thermal energy of its environment[14-17]. In contrast, the efficiency of a laser has to be below 100%[18] since part of the electric pump power has to be converted to heat. In addition, the specific arrangement of the PV cells does not affect the ideal performance of the photonic transformer in the ideal case. Therefore, our analysis focuses on using LED(s) facing laterally connected PV cells in series. Using a detailed balance analysis, we show that the efficiency of a photonic transformer can approach unity. The key to such high efficiency is to achieve highly efficient optical coupling between the LED and the PV cells. We introduce a monolithic design where such a strong optical coupling between the LED and the PV cell can be achieved. We also construct and perform experiments on a photonic transformer consisting of off-the-shelf components. The agreement between experiments and theory provides a validation of our theoretical formalism. Moreover, although these off-the-shelf components are not expected to deliver high efficiency, we are able to achieve a large voltage conversion ratio (> 40) that significantly exceeds what have been previously demonstrated in existing experiments on photonic transformers (< 10)[10,11].

To derive the theoretical performance limit of our photonic transformer, we perform an analysis for the setup depicted in Fig. 1c with one LED facing $N$ identical PV cells. For planar LED and PV cells, when $N$ is large, we can reasonably assume the view factors from the LED to each PV cell, $f_{\text{LED}\to\text{PV}}$, are the same. In this case, each PV cell receives the same amount of illumination from the LED, and therefore generates the same voltage. We denote the voltage (current) of the LED and each PV cell as $V_{\text{LED}}$ ($I_{\text{LED}}$) and $V_{\text{PV}}$ ($I_{\text{PV}}$), respectively. Therefore, the input and output voltages are respectively $V_{\text{IN}} = V_{\text{LED}}$ and $V_{\text{OUT}} = NV_{\text{PV}}$, when series resistance is negligible. The input and output currents are respectively $I_{\text{IN}} = I_{\text{LED}}$ and $I_{\text{OUT}} = -I_{\text{PV}}$. The voltage conversion ratio ($W$) can be expressed as

$$W = \frac{V_{\text{OUT}}}{V_{\text{IN}}} = N\frac{V_{\text{PV}}}{V_{\text{LED}}} \tag{1}$$

And the power efficiency ($\eta$) can be obtained as

$$\eta = \frac{P_{\text{OUT}}}{P_{\text{IN}}} = \frac{V_{\text{OUT}}I_{\text{OUT}}}{V_{\text{IN}}I_{\text{IN}}} = N\frac{-I_{\text{PV}}V_{\text{PV}}}{I_{\text{LED}}V_{\text{LED}}} \tag{2}$$

Equation (1) indicates that the conversion ratio depends on how much voltage each PV cell can recover from the photon flux emitted by the LED. The power efficiency is a product of the voltage conversion ratio and the current conversion ratio ($-I_{\text{PV}}/I_{\text{LED}}$). The voltage and current of the LED and each PV cell can be obtained from the detailed balance relations[19,20]:

$$\left[F_{\text{amb}\to\text{LED}} - F_{\text{LED}\to\text{amb}}\right] + N\left[F_{\text{PV}\to\text{LED}} - F_{\text{LED}\to\text{PV}}\right] - R_{\text{LED}}(V_{\text{LED}}) + \frac{I_{\text{LED}}}{q} = 0 \tag{3}$$

and

$$[F_{\text{amb}\to\text{PV}} - F_{\text{PV}\to\text{amb}}] + [F_{\text{LED}\to\text{PV}} - F_{\text{PV}\to\text{LED}}] - R_{\text{PV}}(V_{\text{PV}}) + \frac{I_{\text{PV}}}{q} = 0 \qquad (4)$$

where $F_{a\to b}$ with $a, b$ = LED, PV, and amb (the ambient), is the photon flux emitted from object $a$ and absorbed by object $b$, $q$ is the elementary charge, $R$ is the total rate of nonradiative recombination. We use a sign convention such that a positive current flows from the p to the n region internally in each diode. Therefore, for normal operation of the photonic transformer, $I_{\text{LED}} \geq 0$ and $I_{\text{PV}} \leq 0$. The ambient term includes all objects beyond the active regions of the LED and the PV, so the photon flux absorbed by the ambient represents the photon leakage from the converter, which should be minimized. The photon flux emitted by the ambient is much smaller compared to the photon flux emitted by the LED, and therefore can be neglected.

For the setup up in the above panel of Fig. 1c, we first consider an ideal scenario where the LED and PV cells have the same bandgap with unity external quantum efficiency (EQE) above the bandgap, which implies that the nonradiative terms in Eqs. (3) and (4) are zero. In this case, we can model the emitted photon flux from object $a$ and absorbed by object $b$ as

$$F_{a\to b}(V_a) = A_a f_{a\to b} \int_{\omega_g}^{\infty} \frac{\omega^2}{4\pi^2 c^2} \frac{1}{\exp\left(\frac{\hbar\omega - qV_a}{kT}\right) - 1} d\omega \qquad (5)$$

In Eq. (5), $a$ and $b$ denote the LED, PV, $\omega$ is the angular frequency, $\omega_g$ is the bandgap frequency, $c$ is the speed of light in vacuum, $k$ is the Boltzmann constant, $V_a$ is the voltage applied on object $a$, and $qV_a$ thus corresponds to the chemical potential of emitted photon[21]. $T$ is the temperature of the diode, and throughout the paper in the numerical calculations, we assume all objects are at room temperature with $T$ = 300 K. $A_a$ is the emitting surface area. $f_{a\to b}$ is the view factor from object $a$ to object $b$. If we further assume no photon leakage, then the terms in Eqs. (3) and (4) involving the ambient become zero, and we have $A_{\text{LED}} = NA_{\text{PV}}$, $f_{\text{LED}\to\text{PV}} = 1/N$, and $f_{\text{PV}\to\text{LED}} = 1$. One therefore could obtain $-NI_{\text{PV}} = I_{\text{LED}}$ based on these equations. Therefore, $W = N\eta$ based on Eqs. (1) and (2), and the maximum efficiency point is when $F_{\text{LED}\rightarrow\text{PV}} = F_{\text{PV}\rightarrow\text{LED}}$ based on Eqs. (3) and (4), i.e., $V_{\text{PV}} = V_{\text{LED}}$ based on Eq. (5). In other words, each PV cell at open circuit condition will fully recover the voltage of the LED. In this case, the series-connected PV cell array is in open circuit condition (i.e., $P_{\text{OUT}} = 0$) and outputs a boosted voltage that is $N$ times the input voltage according to Eq. (1). Therefore, the conversion ratio of the proposed photonic transformer depends on the number of PV cells. This dependence allows one to get in principle any desired high conversion ratio by selecting $N$.

The operation principle of photonic transformers is fundamentally different from that of conventional transformers or switch-mode converters. The emission process in the LED and absorption process in the PV cell side are quantum processes[22]. In contrast, in the traditional transformers (Fig. 1a) or switch-mode converters (Fig. 1b), the power exchange process can be described entirely classically.

Now we analyze the expected performance of an actual photonic transformer. For practical use, with a finite load resistance, the operation point of the PV cells is shifted away from the open-circuit condition in order to have non-zero output power. Hence the voltage recovery in general will not be complete. The voltage recovery will also be subject to penalties from the nonradiative processes and series resistances in both the LED and PV cells. Here, we show that the photonic transformer can still have excellent performance even under these considerations.

We first consider the effect of nonradiative processes. As an example, we choose GaN (bandgap energy $\hbar\omega_g = 3.45$ eV) as the active region material for both the LED and the PV cell since a high-performance GaN-based LED with 95% internal quantum efficiency has been experimentally demonstrated[23]. With the nonradiative recombination included, our computations (See Methods) show that the GaN PV cell can recover over 97% of the voltage of the GaN LED with a power conversion efficiency of over 85% (blue curves in Fig. 1d). The main reason for the imperfect power conversion is the imperfect EQE of the LED (90.3%), which limits the current conversion ratio to about 90%.

We next consider the effect of photonic exchange between the LED and the PV cell. For our device, it is critical to achieving efficient optical coupling between the active regions of the LED and the PV cell. This is similar to a conventional transformer in which one uses a ferromagnetic core to reduce leakage of the magnetic energy. In the case shown as blue curves in Fig. 1d, the LED and the PV cell are separated by an air spacer layer in the far-field regime with a thickness that is much larger than the emission wavelength (365 nm). In this case, only the propagating-wave channels in air with an in-plane wavevector $\beta < \omega/c$ are utilized for light extraction[24] from the front emitting surface. Enhancement of coupling can be achieved if in addition one can utilize channels with $\beta > \omega/c$. A standard approach is to operate in the near-field regime where one reduces the thickness of the air spacer layer to be much smaller than the emission wavelength, so that channels with $\beta > \omega/c$, which are evanescent in air[25,26], can contribute. Maintaining such small thickness in the air separation layer over a large area, however, represents a significant experimental challenge[27-33]. Instead, here we propose to use an Al$_{.5}$Ga$_{.5}$N layer as the spacer layer between the LED and the PV cell. Al$_{.5}$Ga$_{.5}$N is essentially transparent in the emission wavelength range of GaN and has a refractive index $n_{AlGaN}$ similar to that of GaN. As detailed in the Methods section, in doing so, the channels with $\omega/c < \beta < n_{AlGaN}\,\omega/c$ are propagating in the AlGaN layer. With an AlGaN spacer layer thickness of 1 μm, the light extraction efficiency of the LED is significantly improved such that the external quantum efficiency of the LED improves to 98.7%, as compared with that of 90.3% for the far-field case with air spacer. The external quantum efficiency of the PV cell is also greatly enhanced because of the index-matching AlGaN layer. The improvement in external quantum efficiency of the LED and the PV cells results in enhanced voltage conversion ratio and conversion efficiency as indicated by the orange curves in Fig. 1d. Importantly, the enhancement here is not a near-field effect. The external quantum efficiency of the LED changes by less than 1% if one reduces the AlGaN spacer thickness from 1 μm to 10 nm.

Thus, the thickness of the spacer can be larger than the wavelength without affecting the optical coupling. This is important in practice since the spacer layer also needs to provide electrical insulation between the LED and the PV cell.

In Fig. 1e, we show the wall-plug efficiency, i.e., the ratio of the emitted optical power and electrical power consumed by the LED, for three cases. The yellow curve corresponds to the theoretical upper bound in the radiative limit case where the nonradiative recombination rates are zero. As mentioned before, the wall-plug efficiency of the LED in the ideal case theoretically can exceed 100%[12,13]. The orange curve and the blue curves are the AlGaN spacer case and the air spacer case. The wall-plug efficiency is greatly enhanced in the AlGaN spacer case, approaching the radiative limit, as compared with the far-field air spacer case. For example, the peak value of the wall-plug efficiency is enhanced from 90.6% to 99.3%, resulting in an excellent overall conversation efficiency near 97%. In Methods, we further show that the use of the AlGaN spacer layer can greatly improve the current density and the power density of the photonic transformer compared to the air spacer case. The enhanced maximum power density can reach 10 kW/cm$^2$ for an input voltage near 3.4 V, indicating that 1 Watt electrical power can be delivered in a footprint ~0.01 mm$^2$. By comparison, a high-efficiency switch-mode converter may have a volumetric power density[34] of 1 Watt per 10-100 mm$^3$, a much larger footprint compared to photonic transformers. The high power density may enable the use of the GaN photonic transformer in high power applications.

We now explore the penalties from the series resistance. We evaluate the case with the AlGaN spacer layer as considered above which yields superior performance. In the presence of series resistances of the LED and the PV, denoted as $R_{s,\text{LED}}$ and $R_{s,\text{PV}}$ respectively, the input and output voltage can be related based on the circuit diagram in Fig. 1c as

$$V_{\text{IN}} = V_{\text{LED}} + I_{\text{LED}} \frac{R_{s,\text{LED}}}{A_{\text{LED}}} \tag{6}$$

and

$$V_{\text{OUT}} = N\left(V_{\text{PV}} + I_{\text{PV}}\frac{R_{\text{S,PV}}}{A_{\text{PV}}}\right) \qquad (7)$$

For GaN LEDs, a series resistance as small as $1\,\text{m}\Omega\cdot\text{cm}^2$ has been demonstrated experimentally[23], and GaN tunnel junctions with series resistance as low as $0.01\,\text{m}\Omega\cdot\text{cm}^2$ have been demonstrated as well[35]. In Fig. 1f we show the modeled photonic transformer performance for series resistances of 1, 0.1, and $0.01\,\text{m}\Omega\cdot\text{cm}^2$. In general, the performance degrades as the series resistance goes up. For high input voltages above 3.2 V, the series resistance penalty becomes the dominant limitation on the voltage conversion ratio and further increasing the input voltage leads to diminished performance. However, the voltage ratio at the maximum efficiency point can still reach over 90 (for $N = 100$) for all three cases, and the peak efficiency can exceed 90% when $R_s = 0.01\,\text{m}\Omega\cdot\text{cm}^2$, indicating that the excellent performance persists even in the presence of realistic series resistance. We note that, in theory, conventional switching converters can also have arbitrary voltage conversion ratio by controlling the duty cycle. However, parasitic losses in the circuit[1] typically place a severe limit on the useful range of conversion ratios to the order of ten[36,37]. In contrast, the conversion ratio of our photonic transformer is not subject to such a limit. The high conversion ratio and efficiency indicate the great potential for photonic transformers to outperform conventional switching converters[37].

Based on the above analysis, we propose a monolithic solid-state conceptual device design illustrated in Fig. 1g. Separating the GaN LED layers and PV cell layers is the index-matching AlGaN layer that provides both the necessary optical coupling and electrical insulation. Since the combination of LEDs and PV cells can be readily miniaturized and monolithically integrated on a single die, the photonic transformer can be made with a far smaller footprint than those of existing switch-mode DC converters, which require a large, often off-chip, amount of space. The design concept shown in Fig. 1g uses one LED. In this case, some emitting areas of the LED are not directly facing PV cells and therefore contribute to a photon loss. On the other hand, as discussed in the Methods, in the ideal case, photonic transformers using either one LED or multiple LEDs connected in parallel have identical theoretical performance, provided that the total emitting area of the LED is the same. Therefore, in the conceptual monolithic design, one can employ one LED for each PV cell to achieve a better optical coupling between the LED and the PV cell, and doing so may also allow better control over the voltage conversion ratio.

In addition to its high performance in terms of conversion ratio and efficiency, the use of a steady photon flux enables the photonic transformer to produce a ripple-free output voltage. Here, therefore, the contributions from the photonic transformer to the output voltage fluctuations are primarily the photon shot noise and the thermal noise from the series resistance, both of which are fundamental in nature. These noises have broad frequency spectra. But, even when integrated over the entire frequency bandwidth, the power in such noises are still small in comparison with typical thermal noise power associated with the load resistance (Methods). This contrasts with the switch-mode converters where the output voltage ripple as well as the accompanying EMI are an unavoidable result of switching.

To validate the theory above on photonic transformer we construct a circuit prototype using commercially available off-the-shelf LEDs and PV cells. We use multiple LEDs connected in parallel, as shown in the printed circuit board (PCB) design in Fig. 2a. Here we use the same number ($N$) of the LEDs and the PV cells. We choose $N = 100$ to show the high conversion ratio that the photonic approach enables. We use GaAs LEDs (Osram SFH4253-Z) and Si PV cells (Osram BPW 34 S-Z) to ensure reasonable spectral overlap between the LED and the PV cell. We note that these choices are only for demonstration purposes and are far from the optimized devices discussed above. Figure 2b shows the LED and PV cell circuits (on a custom-designed circuit board) and the assembled prototype consisting of the LED board mounted to a corresponding PV cell board. (See Methods for the circuit construction.) To characterize the photonic transformer prototype, we connect the LED board to a DC power supply (Keysight E36312A) and measure the current-voltage (*I-V*) curve of the PV cell board using a source meter (Keithley 2636B) at different input voltage levels. We obtain the maximum efficiency of the transformer at a given input voltage level by locating the maximum power point of the PV cell array on the measured *I-V* curve (See Extended Data for the measured *I-V* data). In Fig. 2c we show the voltage ratio at the maximum efficiency point and the corresponding efficiency of the transformer for different input voltages. The efficiency peaks at the input voltage of $V_{IN} = 1.39 \text{ V}$. Further increasing the input voltage leads to a decrease in efficiency due to the series resistance of the LEDs and the PV cells, as discussed earlier in the theoretical calculation shown in Fig. 1f. At the operation point for peak efficiency (5.77%), we obtain a voltage conversion ratio 31.2, a clear demonstration of the DC voltage conversion functionality of our photonic transformer. The ratio between the open-circuit voltage and the input voltage is 40.9. Therefore, one could tune the operation point to obtain an even higher voltage conversion ratio. While the efficiency demonstrated in our experiments is relatively modest (5.77%), our demonstrated voltage conversion ratio (>40) significantly exceeds what has been previously demonstrated in existing experiments on photonic transformers (< 10)[10,11]. Based on the same theoretical model that has been used to analyze the GaN transformer above, we analyze the proof-of-concept photonic transformer circuit as detailed in Methods, and obtain the predicted maximum efficiency of the transformer and the corresponding voltage ratios at given input power levels. The predictions are shown in Fig. 2c as continuous curves, which agree well with the experimentally measured values. This agreement provides validation of our theoretical model.

The use of steady continuous-wave (CW) photon transport in the photonic transformer enables its low-noise operation despite the high voltage conversion ratio, when compared with the switching action in the switch-mode converter. As mentioned, the switching action in conventional switch-mode converters produces undesirable effects as switching currents not only generate noise in the circuit in the form of ripples in the output voltage level but also emit outgoing radiation contributing to EMI, which can corrupt the operation of sensitive circuits nearby or interfere with signal transmission[7]. To illustrate this point, we measure and compare the electromagnetic noise generated by the conventional switch-mode converter and the photonic transformer using the setup shown in Fig. 3a. The setup consists of an oscilloscope to monitor the output voltage fluctuations and a field probe connected to a spectrum analyzer to monitor the emitted electromagnetic fields (Methods). For the switch-mode converter, we sample two commercial DC voltage up-converters (MCP1640EV from Microchip and LT3482EUD from Analog Devices, hereafter referred to as converters 1 and 2, respectively). Both circuits use a boost configuration (see Fig. 3b for the circuit diagram) with converter 2 featuring a voltage doubler stage at the output to further increase the conversion ratio (27 compared to 3.3 for converter 1). For converter 1, the output contains voltage ripples (Fig. 3d) at its switching frequency of 500 kHz from the underlying charging and discharging cycles. Further, the circuit produces outgoing radiation at the switching frequency and its harmonics which shows up prominently as sharp peaks in the radiation spectrum (Fig. 3g). For converter 2 which uses a higher switching frequency (1.1 MHz) the output voltage ripples are greatly reduced (Fig. 3e) due to the increased effectiveness of the output capacitor filter at higher frequency and higher load impedance. However, even though converter 2 is designed to output voltage with extremely low noise (based on the datasheet of the product) as observed, its radiation spectrum still features a prominent set of peaks at the switching frequency and its harmonics (Fig. 3h). In contrast, our photonic transformer exhibits intrinsically low noise operation (smaller than the background level), both in terms of the output voltage fluctuations and radiative electromagnetic noise. The photonic circuit (Fig. 3c) is free of any internal switching mechanism. As a result, the output voltage waveform (Fig. 3f) exhibits no noticeable ripples even without using an output capacitor, and the field measurements (Fig. 3i) show no detectable emission above the background level. Here, the contributions to the voltage noise power at load are as follows: 21 nV$^2$/Hz from shot noise in PV cells and 55 nV$^2$/Hz from the series resistance of all PV cells; these are small compared with the room temperature thermal noise power of the load resistance ($R_L$) in isolation, $4kTR_L$ = 1100 nV$^2$/Hz (Methods). We note that under our test conditions, the efficiencies of the commercial electronic converter 1 and 2 are respectively 78% and 46%, which are higher compared to the proof-of-concept photonic transformer prototype we demonstrated. In Methods, we conduct an analysis of the nonidealities in the photonic transformer prototype and show that the efficiency of even the far from ideal far-field GaAs-Si photonic transformer with an air spacer can be improved to over 40% using experimentally demonstrated or routinely achieved values for the limiting parasitic loss factors.

As final remarks, we note that the monolithic photonic transformer can also operate as a step-down DC transformer if the LEDs are in series and PV cells are in parallel. The monolithic photonic transformer is highly scalable and can be easily integrated on chip[38]. The conversion ratio or the output voltage can be modified by controlling how many PV cells are connected in the circuit as well as the circuit topology. Other high-quality semiconductors[13] may be used depending on the application and input voltage range. Our photonic transformer also inherently provides electrical isolation between the input and output, protecting the load from destructive or hazardous electric shocks. While the initial application of the photonic transformer concept is likely in low power electronic circuits, one may envision that this concept can be scaled up to a power level relevant for electric power network. The proposed photonic transformer highlights the significant potential for using photons as the intermediate energy carrier in power conversion applications.

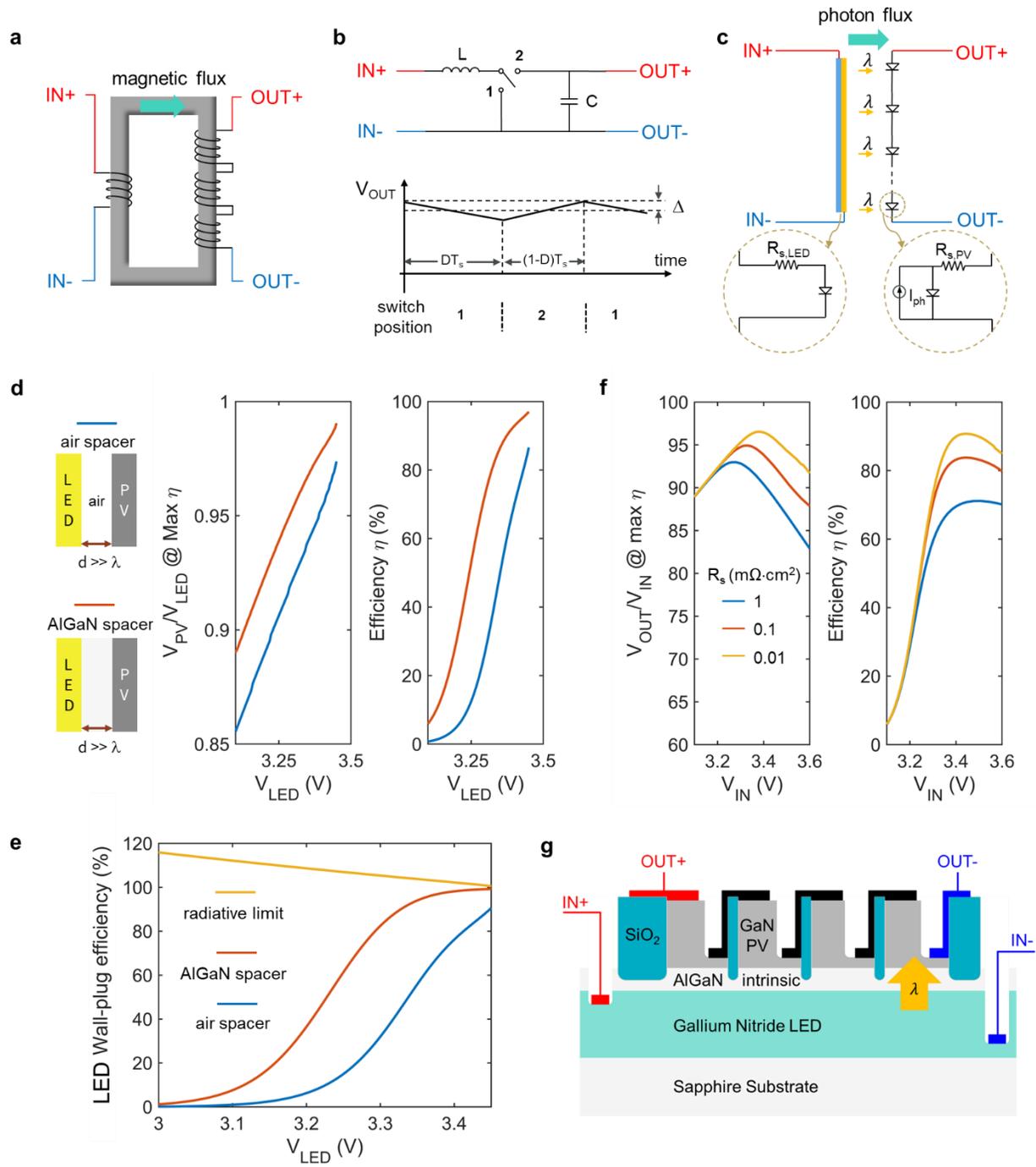

**Figure 1: Designs of an AC transformer, a step-up switching DC converter, and our proposed DC photonic transformer. a**, Schematic of a conventional transformer. Input (AC) voltage on the primary winding generates an alternating magnetic flux in the magnetic core (grey rectangular ring), which induces a scaled version of the voltage on the secondary winding. The voltage

conversion ratio is determined by the winding ratio between the two coils. **b**, Schematic of a typical switch-mode step-up (boost) converter circuit and its steady-state output voltage as a function of time under linear-ripple approximation[1]. The circuit operates by temporally alternating the switch position between 1 and 2 resulting in the charging and discharging of the energy storage elements. $D$: duty cycle. $T_s$: switching period. $\Delta$: output voltage ripple magnitude. **c**, Schematic of the proposed DC photonic transformer. The photon flux emitted by the LED and received by the PV cell acts as a current source for the external load of the PV cell. The inset shows the series resistances of the PV cell and LED. **d**, Theoretical performance of the proposed GaN photonic transformer with the nonradiative recombination included (Methods). The light blue lines correspond to the case of a 1-μm thick air spacer layer between the LED and the PV cell, whereas the orange lines are for the case of an $Al_{.5}Ga_{.5}N$ spacer layer with a thickness of 1 μm. Shown in the efficiency plot is the maximum possible efficiency by adjusting the load. **e**, The wall-plug efficiency of the LED in three cases: the radiative limit (i.e., nonradiative recombination rates are zero), the AlGaN spacer case, and the air spacer case. **f**, The performance of GaN photonic transformer in the near field with different series resistance $R_s = R_{s,PV} = R_{s,LED}$. We assume the number of PV cells $N = 100$. **g**, A conceptual monolithic device design of the proposed GaN photonic transformer with AlGaN electrically-isolating and optical-coupling spacer layer.

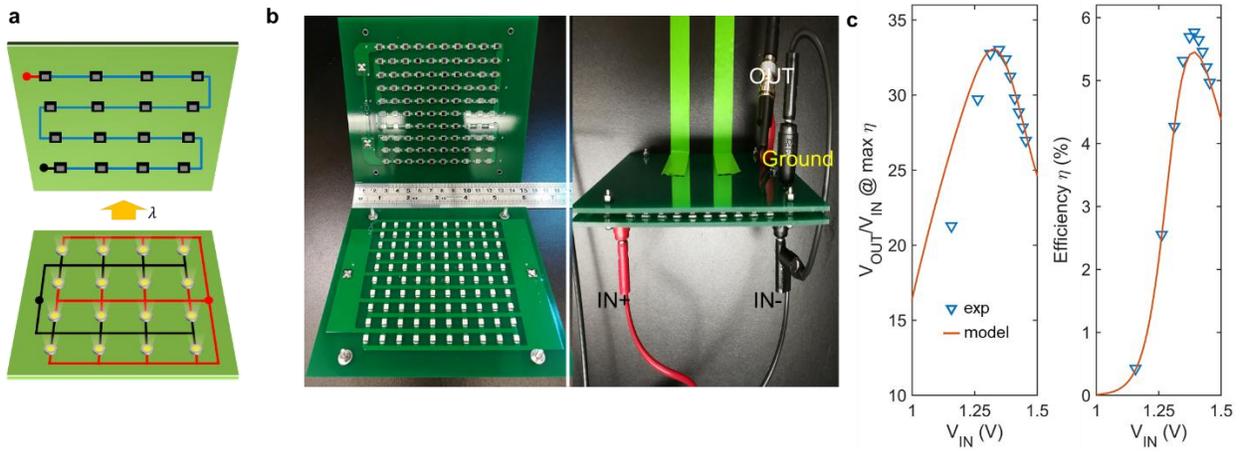

**Figure 2: Proof-of-concept verification of the power conversion functionality of the DC-to-DC photonic transformer. a**, Schematic of a board design with the same number of LEDs (bottom half) and PV cells (top half). **b**, Photos of the two printed circuit boards consisting of LEDs and PV cells (left) and the assembled photonic transformer (right). **c**, Measured (blue triangles) voltage ratio and conversion efficiency. The red curves are predictions from the model.

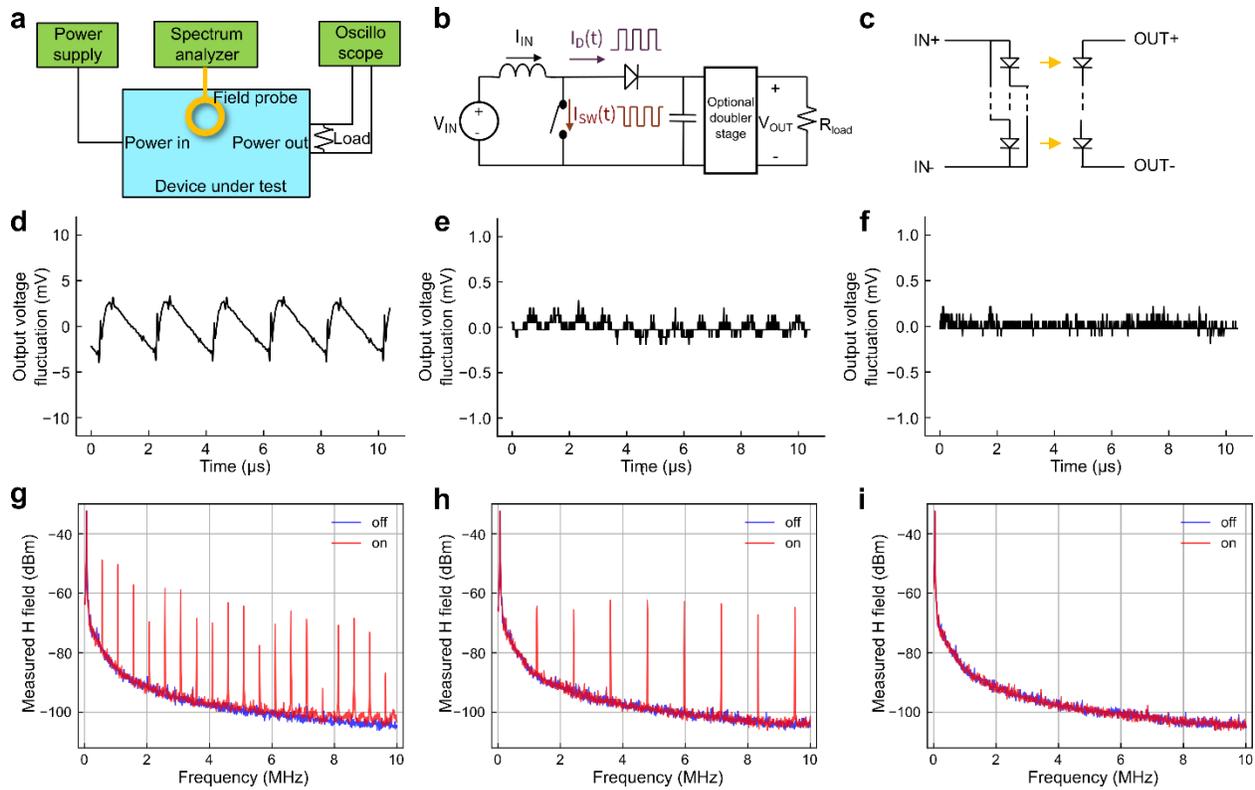

**Figure 3: Experimental measurements of the output noise of commercial switching converters and the photonic transformer. a**, Noise measurement setup. **b**, Switching current flows in a boost converter circuit. **c**, Photonic transformer circuit. **d**-**f**, Output voltage waveforms for (**d**) converter 1, (**e**) converter 2, and (**f**) the photonic transformer. **g-i**, Power spectra of the electromagnetic field emitted from the three devices when the corresponding converter/transformer is power on (red) and off (blue): (**g**) converter 1, (**h**) converter 2, and (**i**) the photonic transformer.

**Methods**

**Fluctuation-dissipation theorem.** In the theoretical description of the GaN photonic transformer, we use the fluctuation-dissipation theorem to model the radiative recombination rate since this approach is applicable for both the near- and far-field scenarios, and one can describe the photon transport process without explicitly defining the radiative recombination coefficient, escape probability, and light extraction efficiency. The computation setups for the GaN photonic transformer are depicted in the Extended Data Figs. 1a and 1b. Here we consider the case where the surfaces of the LED and the $N$ PV cells are parallel, and the schematic shows a unit cell of the whole structure. The above-bandgap emission from a diode is due to a fluctuational current source $\mathbf{j}$ that satisfies[39]

$$\langle j_k(\mathbf{x}',\omega) j_n^*(\mathbf{x}'',\omega') \rangle = \frac{4}{\pi} \omega \varepsilon_0 \, \mathrm{Im}(\varepsilon_e) \Theta(\omega,T,V) \delta_{kn} \delta(\mathbf{x}'-\mathbf{x}'') \delta(\omega-\omega') \tag{M1}$$

where $k$ and $n$ denote the directions of polarization, $\mathbf{x}'$ and $\mathbf{x}''$ are position vectors, $\mathrm{Im}(\varepsilon_e)$ is the imaginary part of the dielectric function, $\varepsilon_0$ is the vacuum permittivity, $\delta$ is the Dirac delta function, and $\Theta$ is the photon energy in a mode at $\omega$

$$\Theta(\omega,T,V) = \frac{\hbar\omega}{\exp\left(\frac{\hbar\omega-qV}{kT}\right)-1} \tag{M2}$$

where $V$ is the bias on the active region and $qV$ acts as the chemical potential of photons[17,21]. Using the formalism of fluctuational electrodynamics, the energy transfer between the LED and the PV cell can be modeled as

$$Q = Q_{\mathrm{LED}\rightarrow\mathrm{PV}}(V_{\mathrm{LED}}) - Q_{\mathrm{PV}\rightarrow\mathrm{LED}}(V_{\mathrm{PV}}) \tag{M3}$$

where

$$Q_{\mathrm{LED}\rightarrow\mathrm{PV}}(V_{\mathrm{LED}}) = \frac{A_{\mathrm{LED}} f_{\mathrm{LED}\rightarrow\mathrm{PV}}}{8\pi^3} \int_{\omega_g}^{\infty} d\omega \iint \xi(\omega,k_x,k_y) \Theta(\omega,V_{\mathrm{LED}}) dk_x dk_y \tag{M4}$$

and

$$Q_{\text{PV}\rightarrow\text{LED}}(V_{\text{PV}}) = \frac{A_{\text{PV}} f_{\text{PV}\rightarrow\text{LED}}}{8\pi^3} \int_{\omega_g}^{\infty} d\omega \iint \xi(\omega, k_x, k_y) \Theta(\omega, V_{\text{PV}}) dk_x dk_y \quad (\text{M5})$$

In the above two equations, $\xi(\omega, k_x, k_y)$ is the energy transmission coefficient summing over two polarizations and has a maximum value of 2. The nonideality from less than 100% efficiency of light extraction from the LED and absorption from the PV cell are reflected in the energy transmission coefficient being less than 2. *A* is the emitting surface area. For the case we considered here $A_{\text{LED}} = NA_{\text{PV}}$, $f_{\text{LED}\rightarrow\text{PV}} = 1/N$, and $f_{\text{PV}\rightarrow\text{LED}} = 1$. The photon flux can be obtained accordingly as

$$F_{\text{LED}\rightarrow\text{PV}}(V_{\text{LED}}) = \frac{A_{\text{LED}} f_{\text{LED}\rightarrow\text{PV}}}{8\pi^3} \int_{\omega_g}^{\infty} d\omega \iint \xi(\omega, k_x, k_y) \frac{\Theta(\omega, V_{\text{LED}})}{\hbar\omega} dk_x dk_y \quad (\text{M6})$$

and

$$F_{\text{PV}\rightarrow\text{LED}}(V_{\text{PV}}) = \frac{A_{\text{PV}} f_{\text{PV}\rightarrow\text{LED}}}{8\pi^3} \int_{\omega_g}^{\infty} d\omega \iint \xi(\omega, k_x, k_y) \frac{\Theta(\omega, V_{\text{PV}})}{\hbar\omega} dk_x dk_y \quad (\text{M7})$$

The nonradiative recombination rates of the LED and PV cell are

$$R_a = A_a \left( C_{n,a} n_a + C_{p,a} p_a \right) \left( n_a p_a - n_{i,a}^2 \right) t_a + \frac{1}{\tau_a} \frac{n_a p_a - n_{i,a}^2}{n_a + p_a + 2n_{i,a}} A_a t_a \quad (\text{M8})$$

In the above equation, *a* = PV cell or LED. The first and the second terms on the right-hand side are the Auger and Shockley-Read-Hall (SRH) recombination rates, respectively. $\tau$ is the bulk SRH lifetime, and $C_p$ and $C_n$ are the Auger recombination coefficients for holes and electrons, respectively. Using Eqs. (M6), (M7), and (M8), together with Eqs. (3) and (4), we obtain a model of the *I-V* curve for both the LED and PV cell. The computation code for the model can be accessed from Ref. [40].

Equations (M4)-(M7) contain integration over $k_x$ and $k_y$. In principle, there is no upper limit for $k_x$ and $k_y$ in such integral. However, in the far-field case with air, the energy transmission coefficient is non-negligible only for $\beta^2 = k_x^2 + k_y^2 \leq \omega^2/c^2$, where $c$ is the speed of light in air. In contrast, in the case with an AlGaN spacer layer, there are significant contributions from regions with $\beta^2 > \omega^2/c^2$. As an illustration, we show the energy transmission coefficient for the far-field case with air and the case with AlGaN spacer layer in the Extended Data Figs. 1c and 1d. We see that in the far-field case with air, only the propagating-wave channels above the light line of air (green dashed line) are active for photon transport. In contrast, for the case with AlGaN layer, the channels between the light lines of GaN and air now dominate the transport between the LED and the PV cell. The additional contribution of these channels greatly enhances the external quantum efficiency of the LED. In Extended Data Fig. 1e, we show the external quantum efficiency (EQE) of the GaN LED in both the far-field case with air, and the case with AlGaN spacer layer. In the computation, we use the optical properties of GaN and AlGaN from Ref. [41] and take into account the bias effect on the imaginary part of the dielectric function using the formula discussed in Ref. [39]. We model nonradiative terms for both the LED and the PV cell using typical values $C_n = C_p = 5\times10^{-32}$ cm$^6$/s and $\tau = 20$ μs [42]. In the far-field case with air, the GaN LED has an equivalent radiative recombination coefficient about $B = 9\times10^{-12}$ cm$^3$/s, similar to the reported typical value[42]. In the case with AlGaN spacer layer, the light extraction efficiency is greatly enhanced, resulting in an enhanced EQE for the LED as the figure shows. We note that, by enhancing the light extraction efficiency with the use of the AlGaN spacer layer, one also increases the radiative recombination coefficient, since photons previously trapped and reabsorbed by the LED (contributing to a reverse current) can now be extracted. Therefore, the use of the AlGaN spacer layer also greatly improves the current density and the power density of the photonic transformer, as shown in the Extended Data Fig. 1f. The enhanced power density may enable the use of the GaN photonic transformer in high power applications.

**Photonic transformer with *N* LEDs and *N* PV cells.** We perform an analysis on the ideal performance of the setup with identical *N* LEDs and *N* PVs. For the circuit shown in Fig. 2a, the LEDs have the same current and the PV cells have the same voltage. We denote the voltage (current) of each LED and PV cell as $V_{LED}$ ($I_{LED}$) and $V_{PV}$ ($I_{PV}$), respectively. Due to the symmetry of the system, the input and output voltages are respectively $V_{IN} = V_{LED}$ and $V_{OUT} = NV_{PV}$, when series resistance is neglected. The input and output currents are respectively $I_{IN} = NI_{LED}$ and $I_{OUT} = -I_{PV}$. Compared to the one LED and $N$ PV cells case discussed in the main text, the only difference is in the formula for $I_{IN}$. In the ideal case, the total current in $N$ LEDs in parallel is equal to the current in one LED, provided that the total emitting area of the $N$ LEDs is the same with the emitting area of the one LED. Therefore, $I_{IN}$ is equivalent for the $N$ LEDs case and one LED case. Thus, the photonic transformer will have the same theoretical performance for the two cases. Practically, using multiple LEDs may assist the optical coupling between the LEDs and PV cells, and help eliminate the series resistance caused by current spreading in large active area LEDs.

**Construction of the photonic transformer prototype.** We design two circuit boards and have them fabricated by PCBWay – one to populate 100 LEDs (Osram SFH4253-Z GaAs LEDs) and the other to house 100 PV cells (Osram BPW 34S-Z Si PIN photodiodes). The LEDs/PV cells are arranged in a 10 × 10 grid with 1 cm pitch in either direction on the corresponding board and routed to realize a parallel (series) connection on the LED (PV) board. Power connections for both boards are made on the reverse side of the boards. The two boards are mounted with LEDs and PV cells facing each other using alignments holes placed at each board corner through which a series of bolts, spacing washers, and nuts are used to maintain LED-to-PV alignment and ensure good optical coupling.

**Measurement of view factor and LED EQE.** The PV cell in general has a less than 100% probability of converting an incident photon into photocurrent. We denote the external quantum efficiency of the PV cell as $\eta_{RES}$ to account for the nonideal response of the PV cell. Extended data Fig. 2 shows the external quantum efficiency of the Si PV cell and the electroluminescent emission spectrum of the GaAs LED. The external quantum efficiency of the PV cell is defined as the ratio between the output electron number flux and the input photon number flux at the short-circuit condition. The spectrum-averaged external quantum efficiency of the PV cell from 725 nm to 925 nm is $\eta_{RES} = 0.897$. The average photon energy of the LED emission spectrum is 1.46 eV (847 nm). Based on the datasheet, the emitted optical power from the LED is 40 mW at $I_{IN} = 70$ mA. With the averaged emitted photon energy, we compute the external quantum efficiency of the LED at this input power level and find $EQE = 39.1\%$.

To measure the photon transfer efficiency from the LED to the PV cell, we build a separate device that has only one LED and one PV cell. We then measure the ratio of the input current of a LED and the short-circuit current of the PV cell as shown in Extended data Fig. 3, from which

$$\frac{I_{OUT}}{I_{IN}} = EQE \times f_{LED \to PV} \times \eta_{RES} \quad (M9)$$

At $I_{in} = 70$ mA, we measure a current ratio of 0.254. Together with the averaged external quantum efficiency of the PV cell, we obtain $f_{LED \to PV} = 0.73$. In the device model, we assume this view factor is the same for every LED and PV cell pair in the transformer prototype. We then measure the *I-V* curves of the LED array as shown in the Extended data Fig. 4a. For each input level, we measure the *I-V* curve of the PV cell array as shown in Extended data Fig. 5. Based on Eq. (M9), we obtain the EQE of the LED array at different input levels as shown in Extended data Fig. 4b from the ratio of output short circuit current and the input current.

**Device model for the photonic transformer prototype.** Since our prototype is a far-field device, we can simplify the model and highlight the important nonidealities. Instead of Eq. (5), we compute the photon flux produced by the LED as

$$F_0 = A_{LED} B_{LED} n_{LED} p_{LED} t_{LED} \quad (M10)$$

In the above equation, *t* is the thickness of the active region of the diode, *A* is the area of the active region, *n* and *p* are the electron and hole concentrations, respectively, and *B* is the radiative recombination coefficient. Due to the refractive index contrast between the LED and air, many of the generated photons will be trapped in the LED and eventually absorbed parasitically by the LED such as in the contacts. Therefore, we introduce a light extraction efficiency ($\eta_{EXT}$) which describes the proportion of photons that can escape from the LED into free space. The imperfect transmission of light from the LED to the active region of the PV cell is captured by the geometric view factor $f_{LED \to PV}$. We lump the internal photon loss in the LED and the PV cell detection photon loss all in the ambient terms in Eqs. (3) and (4). With these parameters, the photon flux terms in Eqs. (3) and (4) can be modeled as

$$F_{\text{LED}\to\text{PV}} = f_{\text{LED}\to\text{PV}}\eta_{\text{EXT}}\eta_{\text{RES}}F_0 \qquad (\text{M11})$$

and

$$F_{\text{LED}\to\text{amb}} = (1-f_{\text{LED}\to\text{PV}})\eta_{\text{EXT}}F_0 + (1-\eta_{\text{EXT}})F_0 + f_{\text{LED}\to\text{PV}}\eta_{\text{EXT}}(1-\eta_{\text{RES}})F_0 \qquad (\text{M12})$$

In Eq. (M12), the first term on the right-hand side is the photon loss directly to the ambient, the second term is the internal photon loss in the LED, and the third term is the photon loss in the incident photon flux that is received but not absorbed by the active region of the PV. Since the photon fluxes emitted by the Si PV cell and the ambient in general are much smaller compared to that from the emission from the GaAs LED with a bias, we neglect the other photon flux terms in Eqs. (3) and (4). The nonradiative terms are the same as Eq. (M8). Substituting Eqs. (M11), (M12), and (M8) into Eqs. (3) and (4), we obtain a model for the *I-V* curves of the LED and PV cell boards. Besides the parameters that are measured (i.e., view factors and LED EQE), we obtain the remaining parameters used in the model by fitting the measured *I-V* curve of the LED and the set of *I-V* curves of the PV cell iteratively using the *fmincon* function provided by MATLAB. We list the obtained parameters in the tables in the Extended data section.

**Possible improvements for the prototype.** From the device model for the prototype, we identify several aspects that could be improved for better performance. These aspects include the series resistance of the PV, the external quantum efficiency for both the LED and the PV, and the optical coupling between the LED and the PV cell. The series resistance in our device ($R_{\text{s,PV}} = 2.33\ \Omega\cdot\text{cm}^2$) can be improved significantly to as small as $0.32\ \Omega\cdot\text{cm}^2$ by optimizing the PV cell design[22]. This improvement especially helps to minimize the loss at high power levels as shown by the orange curve in Extended Data Fig. 7a. Also, one can use higher quality semiconductor materials to improve the radiative efficiency of the LED and the PV. For the LED, both the Auger process and the SRH process are important nonradiative nonidealities since the LED is operating near its peak of quantum efficiency. For the PV cell, the SRH process is the major nonradiative recombination process because its bias at relevant operating conditions is far below its bandgap. With the relevant parameters replaced by the improved numbers reported in the literature[43,44], efficiency and the voltage conversion ratio can be both significantly improved for all input power levels, as shown by the yellow curves in Extended Data Fig. 7. In addition, the optical coupling between the LED and the PV cell can be improved. This includes the light extraction improvement for the LED, which reduces the internal photon loss inside the devices, and increasing the view factor between the LED and PV cell to suppress photon leakage to the environment. We note that the curve showing 'material' improvements also includes the previously mentioned improvement of the series resistance of the PV cell; the 'optical' curve includes both 'material' and series resistance improvements. With all these improvements implemented, the performance can be raised to what indicated by the purple curves in Extended Data Fig. 7.

**Setup for measuring output voltage fluctuations and electromagnetic field emissions.** Three circuits are used for the performance comparison of Fig. 3: two circuits based on commercial switch-mode converter design – Microchip MCP1640EV (converter 1) and Analog Devices LT3482EUD (converter 2), both are evaluation boards containing their respective manufacturer's suggested design and board layout to implement step-up DC-to-DC converters – and the photonic transformer circuit described above. Each circuit undergoes the following measurement procedures: (i) an appropriate load resistance to produce ~50 mW output power is selected and mounted on the circuit output; (ii) input DC voltage supply (Keysight E36312A adjustable DC power supply) is applied; (iii) output voltage level is measured (B&K 2709B multimeter) and output voltage waveform is taken (Lecroy WaveAce 1012 oscilloscope); and (iv) field emission spectrum is taken using magnetic field probe (Beehive Electronics BH100C) connected to a spectrum analyzer (Tek 495P). In the final step, we maintain a 2-cm parallel gap between the field probe and the circuit board; the location of the probe where the spectrum is taken is the one at which maximum field is registered on the spectrum analyzer as measured by the magnitude of the lowest frequency peak, if available. An extra spectrum is taken with power to the circuit under test turned off to provide measurement of the background/instrument noise floor. Measurement parameters for converter 1, 2, and photonic transformer circuit are, respectively, as follows. Load resistance: 220 Ω, 82 kΩ, 68 kΩ; Input voltage: 1.0, 3.0, 1.5 V; Measured output DC voltage: 3.3, 78.9, 58.2 V. Spectrum analyzer settings: 1 kHz resolution bandwidth, auto sweep rate. (See Extended Data Fig. 8 for the photo of measurement setup).

**Noise analysis for the photonic transformer circuit.** We begin by describing noise processes in a photonic transformer consisting of only one LED and one PV cell (Extended Data Fig. 9a). The photon statistics of an LED connected to a voltage source is well characterized by photon shot noise[45,46]. The noise in the PV cell at low injection levels can be considered as a result of independent noise fluctuations, each modeled as shot noise, as shown in Extended Data Fig. 9b where $I_{photo}$ is the photocurrent, $I = I_0 e^{\frac{qV}{\eta kT}}$ is the junction's forward current, and $I_0$ is the saturation current[45,47]. Hence the combined noise is $S_{I_{PV}}(f) = 2q(I_{photo} + I + I_0) \leq 4qI_{photo}$ where the upper bound is reached at the open-circuited operation ($I_{OUT} = 0$). $R_j = \frac{\eta kT/q}{I}$ and $C_j$ are small-signal junction resistance and capacitance with $R_j C_j \approx$ minority carrier lifetime, $\tau$[48,49]. In addition, the series resistance $R_s$ contributes Johnson noise with $S_{I_{Rs}} = 4kT/R_s$. For a photonic transformer with $N$ PV cells, fluctuations from all the PV cells combine to produce noise at the load $S_V(f) = S_{V,PV}(f) + S_{V,Rs}(f)$ where $S_{V,PV}(f) = NS_{I_{PV}}(f)|H_{PV}(f)|^2$ and $S_{V,Rs} = NS_{I_{Rs}}(f)|H_{Rs}(f)|^2$ are the contributions from $S_{I_{PV}}$ and $S_{I_{Rs}}$, respectively, with $H_{PV}(f) = \left(\frac{Z_j}{NZ_j + NR_s + R_L}\right) R_L$ and $H_{Rs}(f) = \left(\frac{R_s}{NZ_j + NR_s + R_L}\right) R_L$ being the transfer functions from their respective individual noise sources to voltage noise on the load, and $Z_j = R_j || \left(\frac{1}{2\pi i f C_j s}\right)$ (Extended Data Fig. 9c). We evaluate these noise contributions for the photonic transformer circuit under the measurement conditions of Fig. 3 with relevant circuit parameters as follow: $N = 100$, $I_{photo} = 9$mA, $R_j = 6\Omega$, $\tau = 92.8$ns, $R_s = 33\Omega$, and $R_L = 68$k$\Omega$. Since $R_L \gg NR_s$, $S_{V,PV}(f)$ reduces to a flat noise power of $4qI_{photo}NR_j^2 = 21$ nV$^2$/Hz up to the cut-off frequency of $f_c = \frac{1}{2\pi\tau} = 1.7$ MHz and $S_{V,Rs}$ reduces to $4kTNR_s = 55$ nV$^2$/Hz (up to the bandwidth of the oscilloscope). These noise contributions from the photonic transformer represent minor contribution when compared with a standalone resistive load of $R_L$ which produces noise power $4kTR_L = 1100$ nV$^2$/Hz at room temperature – this statement applies in general for near-constant output voltage operation which implies $R_L \gg NR_s$. Finally, we note that low-frequency noise (often referred to as "1/f" or flicker noise), which typically shows up in electronics and manifests as fluctuations over long time scale, may contribute to higher noise at low frequency. Such noise has been found to correlate with defects in semiconductor lattice and contacts, and can be reduced with higher quality device preparation[50,51].

**Data availability** The data that support the findings of this study are available from the corresponding author upon reasonable request.

**Acknowledgments** This work was supported by the U. S. Department of Energy Photonics at Thermodynamic Limits Energy Frontier Research Center under Grant DE-SC0019140 (theoretical model), by a U. S. Army Research Office MURI project under Grant W911NF-19-1-0279 (fluctuational electrodynamic calculation), and by a U. S. Department of Defense Vannevar Bush Faculty Fellowship under Grant N00014-17-1-3030 (experiment). B. Zhao acknowledges Dr. David Miller and Dr. Avik Dutt for stimulating discussions.

**Author contribution** B.Z. and S.A. performed the simulation, modeling, and experiment. P.S. assisted the simulations and modeling. All the authors contributed to formulating the analytical model, to analyzing the data, and to writing the manuscript. B.Z. and S.F. initiated the project. S.F. supervised the project.

**Competing interests** The authors declare no competing interests.

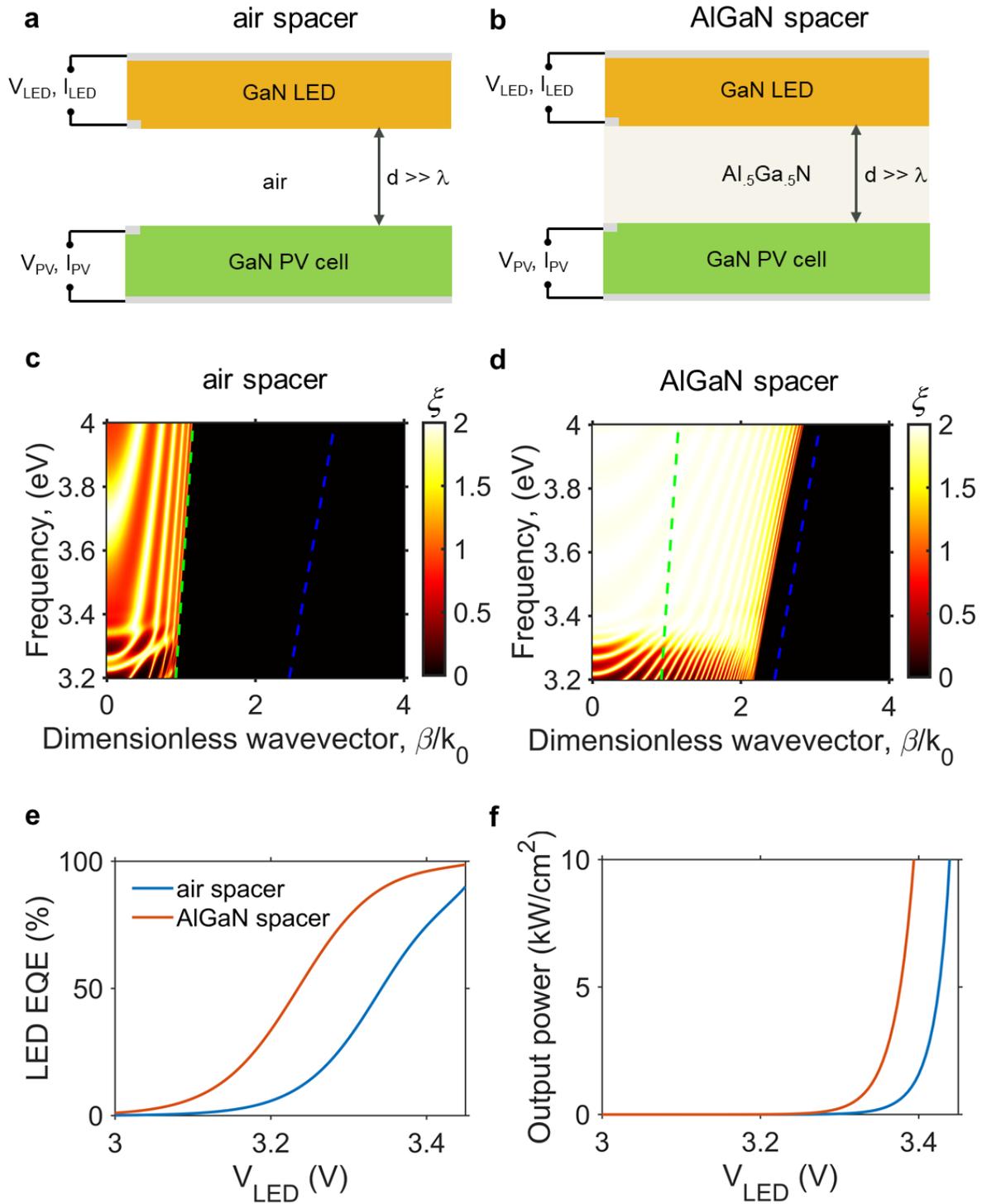

Extended Data Figure 1. **Computation setup of the GaN photonic transformer in Fig. 1d of the main text and the related transport properties in these two cases. a**, The setup for the far-field case with an air spacer layer, where the thickness of the air spacer is $d = 1$ μm. **b**, The AlGaN spacer layer setup where the LED and the PV cell are coupled by a thin $Al_{.5}Ga_{.5}N$ layer, where the thickness of the $Al_{.5}Ga_{.5}N$ spacer layer is also $d = 1$

μm. The thick grey lines on the back of the LED and the PV cell are metallic mirrors for photon recycling and electrical contact purposes. They are modeled as a perfect conductor in the simulation. The thicknesses of the LED and the PV cell are set to be 1 μm for both the far-field and near-field cases. **c** and **d**, Photon transmission coefficients for the cases with air spacer layer and AlGaN spacer layer, respectively. The green dashed line is the light line of the vacuum given by $\beta = \omega/c$, and the blue dashed line is the light line of GaN given by $\beta = n_{\text{GaN}} \omega/c$, where $n_{\text{GaN}} = 2.65$ is the refractive index of GaN near its bandgap. The wavevectors are normalized using the free-space wavevector at the bandgap frequency. **e**, External quantum efficiency (EQE) of the GaN LED for the air spacer and AlGaN spacer cases. In computing the curves, the PV cell is assumed to be in short circuit condition. **f**, Output power density of the GaN PV cell in the air spacer and AlGaN spacer cases at the maximum efficiency (power density) operation point. The EQE and the output power density are greatly enhanced in the AlGaN spacer case as compared to the far-field air spacer case.

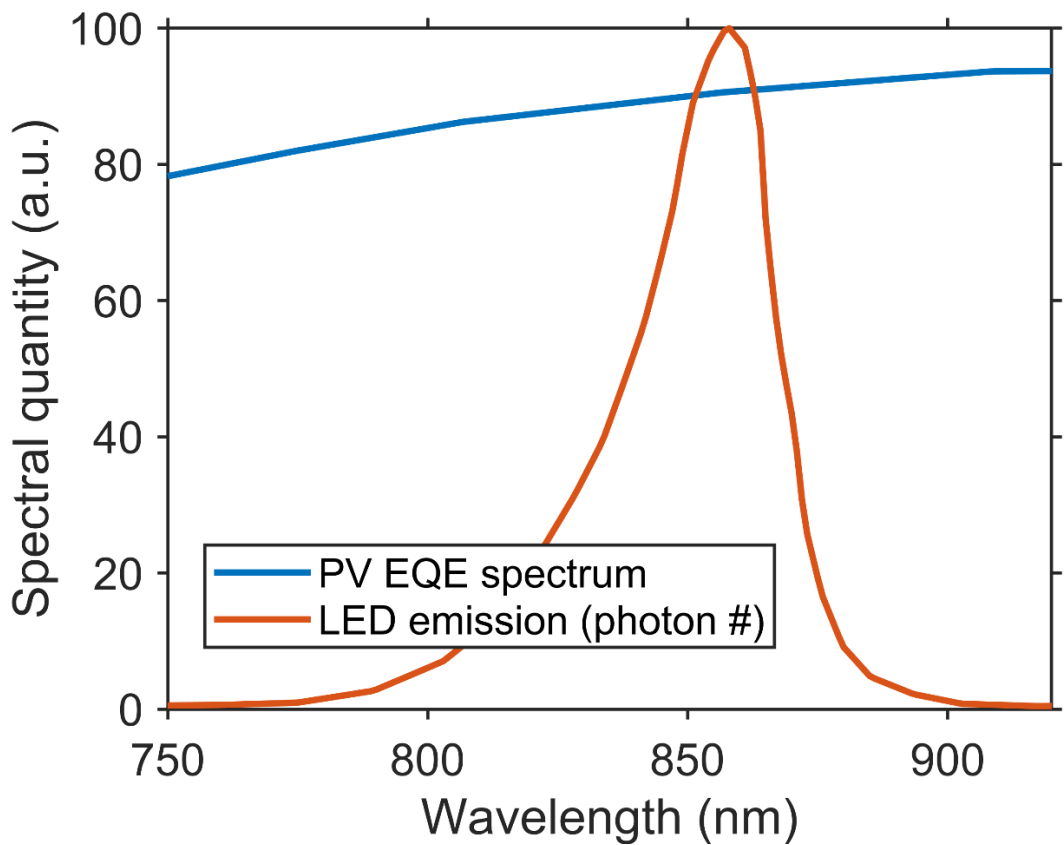

Extended Data Figure 2. **The spectrum of photon flux from the electroluminescence of the GaAs LED and the external quantum efficiency spectrum of the Si PV cell.** Obtained from the datasheets of the devices used in our experiments.

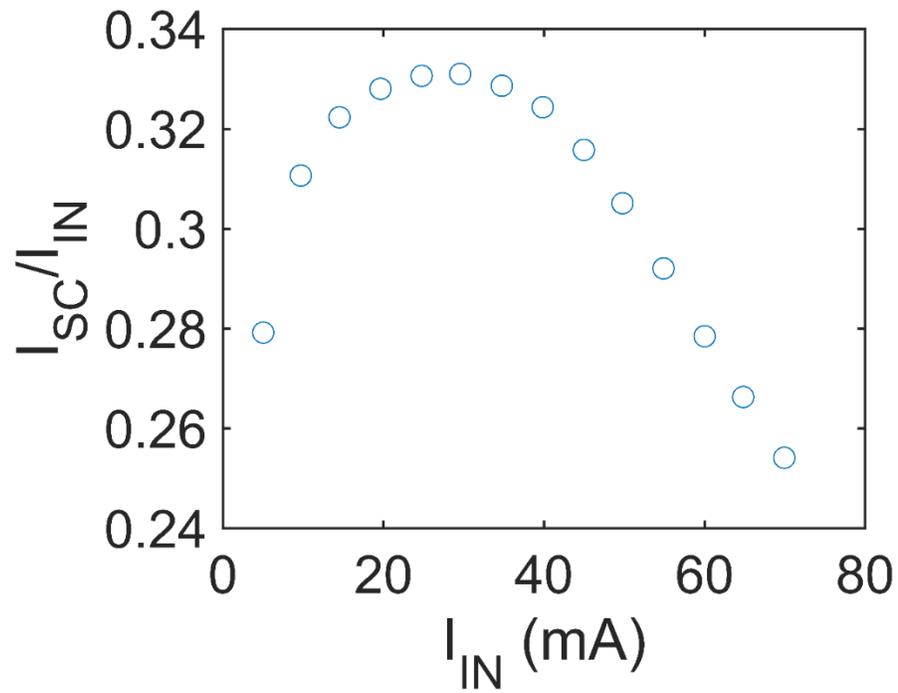

Extended Data Figure 3. **Measurement of a device consisting of one PV cell facing one LED**. Plotted here is the ratio of the short circuit current of one PV cell to the input current of one LED, as a function of the input current of the LED.

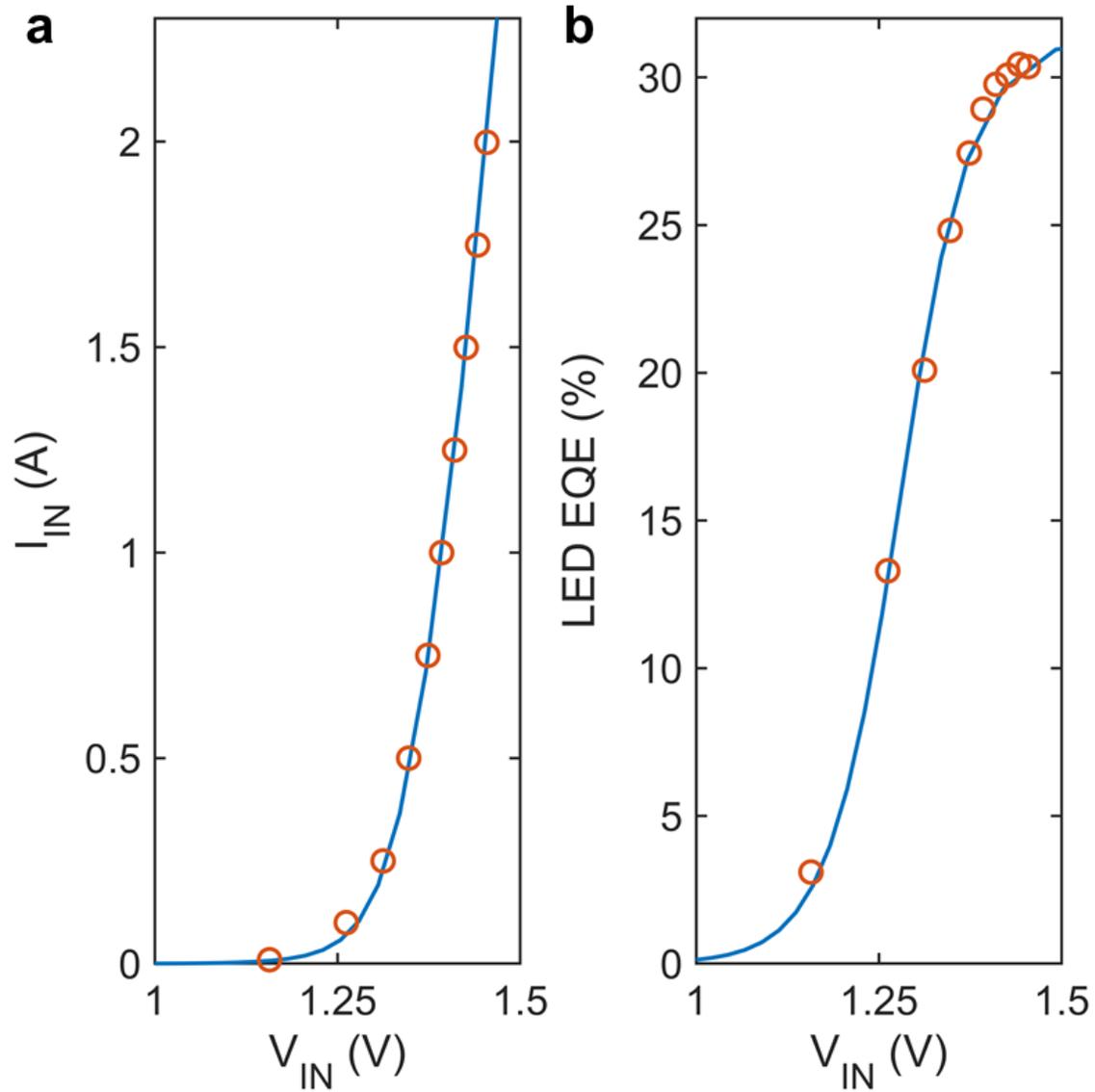

Extended Data Figure 4. **Measured (round dots) and fitted *I-V* curve and external quantum efficiency (EQE) curve of the LED board.** The parameters obtained from numerical fitting are listed in **Extended Data Table 1.**

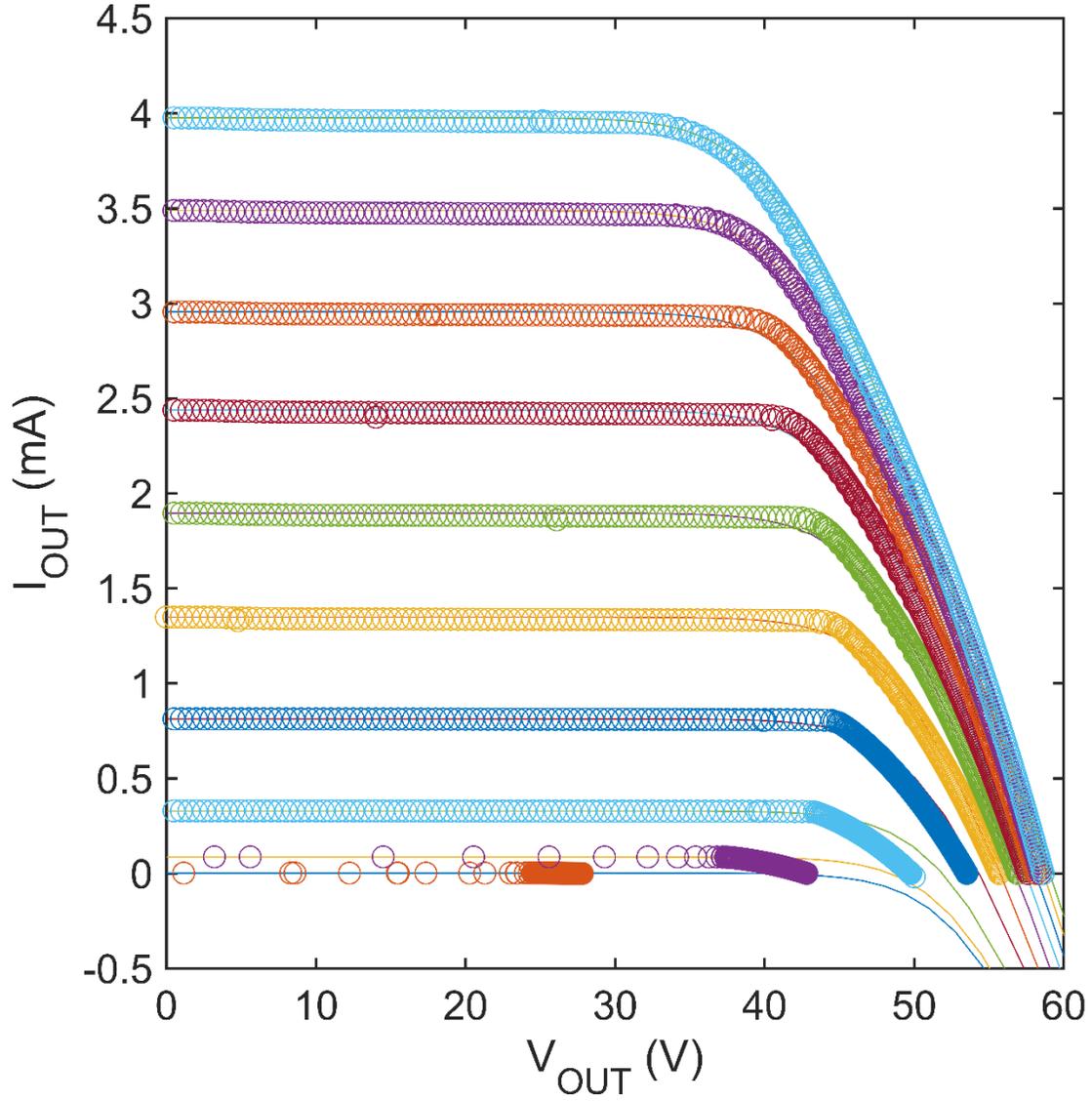

Extended Data Figure 5. **Measured (round dots) and fitted *I-V* curves (continuous lines) of the PV cell array at different input power levels.** The parameters obtained from numerical fitting are listed in **Extended Data Table 1.**

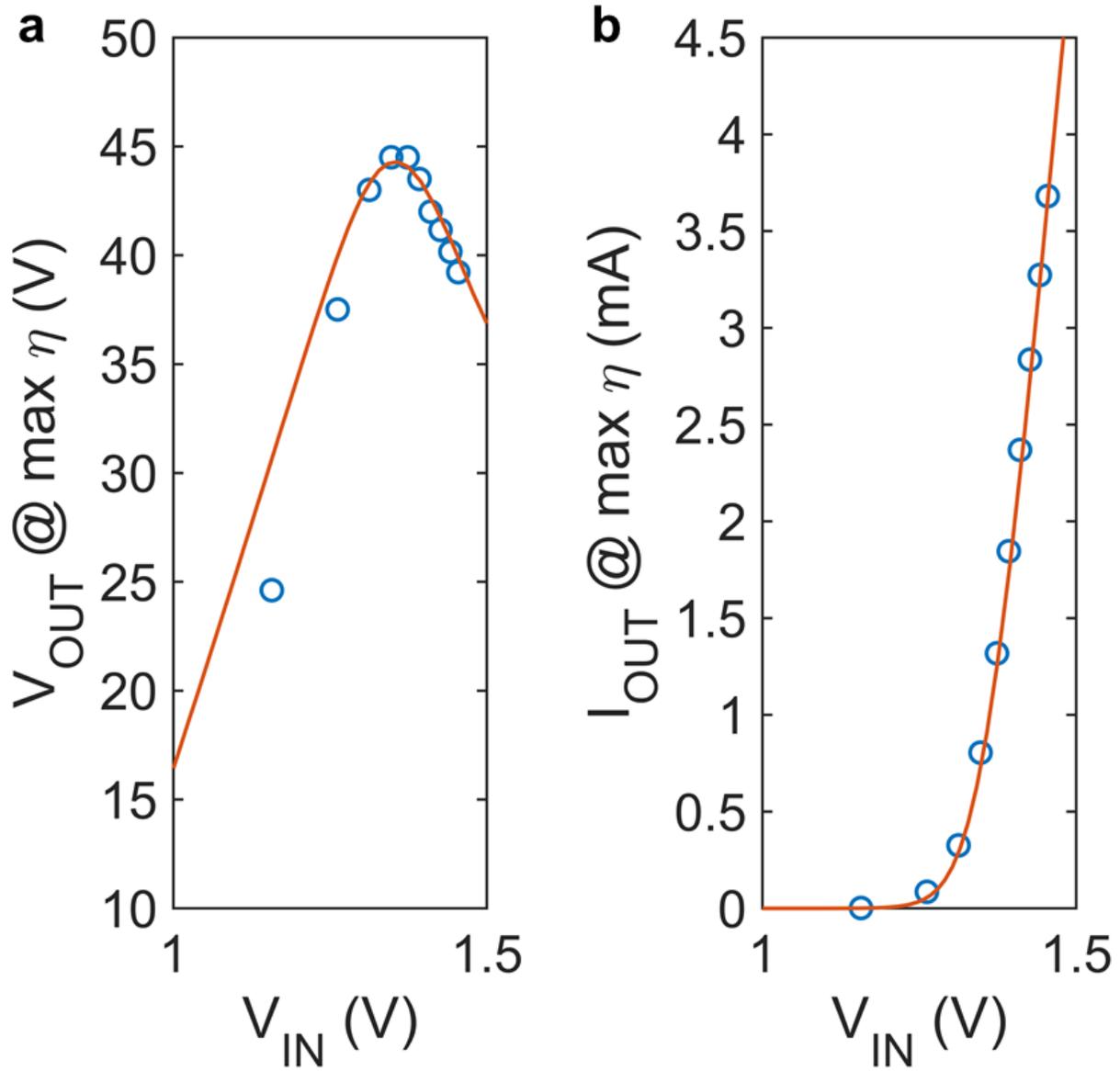

Extended Data Figure 6. **Measured (round dots) and model predictions (continuous lines) of the output voltage and current at the maximum efficiency point.**

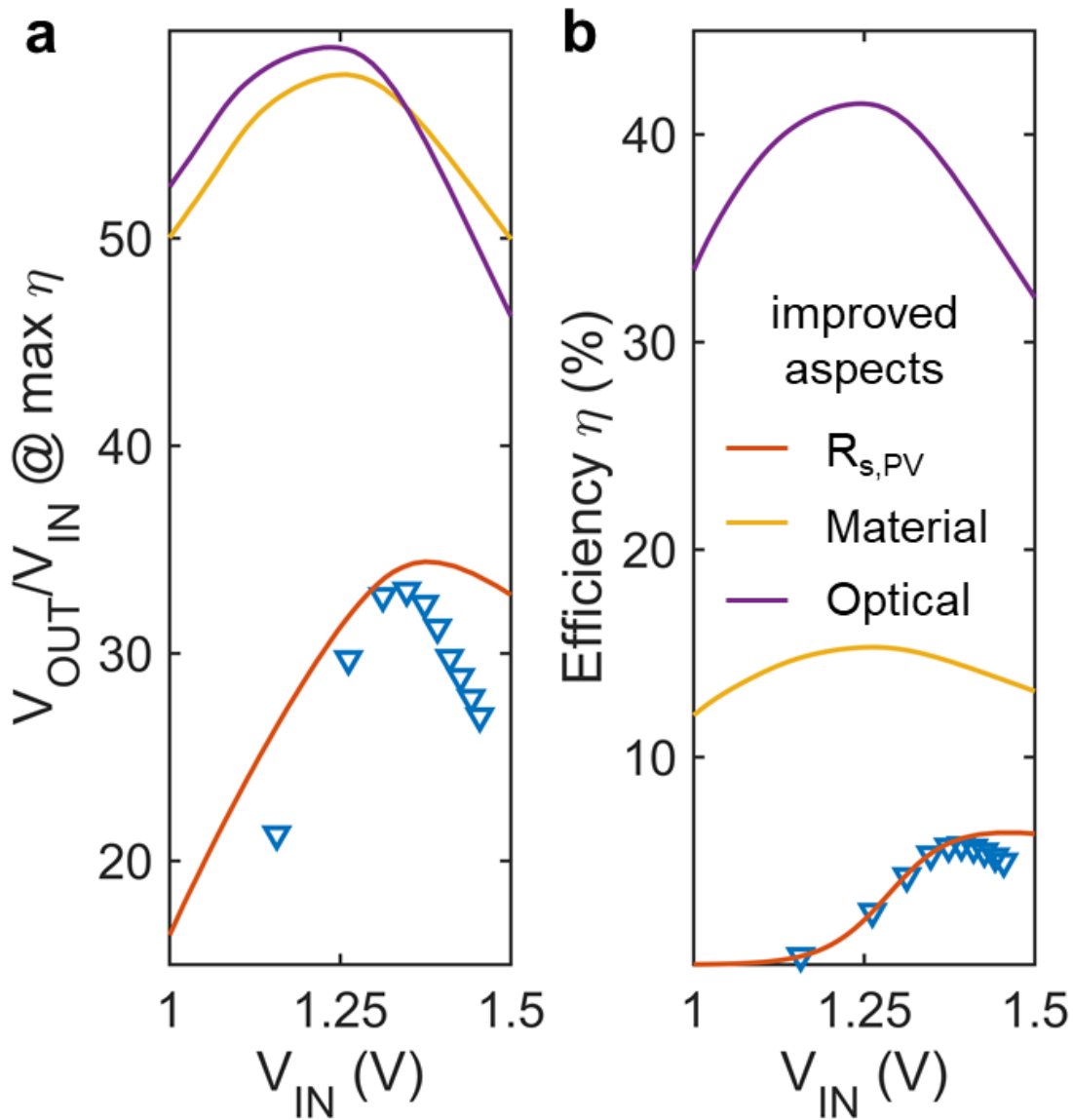

Extended Data Figure 7. **Possible improvements on the photonic transformer prototype.** Efficiency and voltage ratio improvement when a particular aspect of the prototype improves. The triangle data points are the measured data. In obtaining the curves in red, we improve the series resistance of the PV cells to 0.32 Ω·cm$^2$. We further improve the radiative and nonradiative recombination terms for both the LED and the PV cell to obtain the orange curves. For the purple curves, in addition to all the previous improvements, we set the light extraction efficiency to be 80% and the view factor between the LED and the PV cell to be 1. We list the improved parameters in **Extended Data Table 2**.

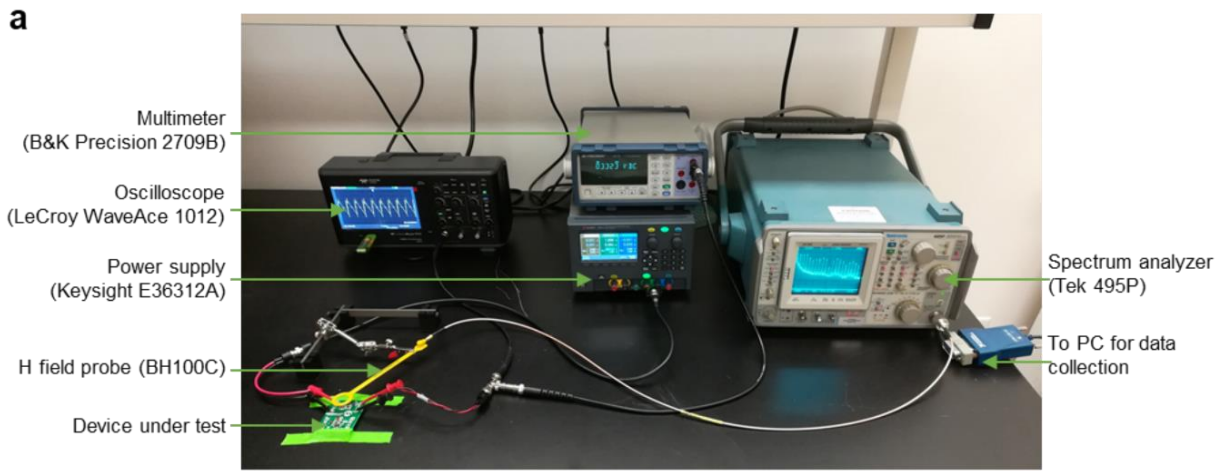
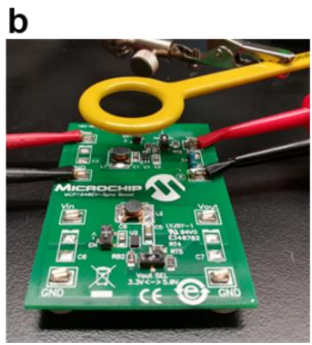
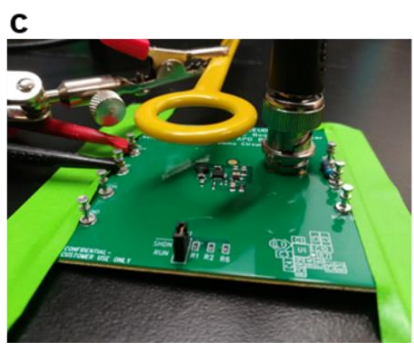
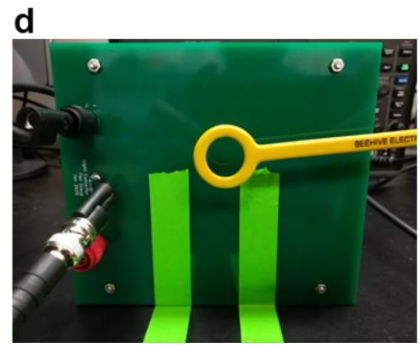

Extended Data Figure 8. **Setup for measuring output voltage fluctuations and electromagnetic field emissions. a**, Measurement setup. **b-d**, Close-up photos of the circuits under test: converter 1 (b), 2 (c), and photonic transformer circuit (d).

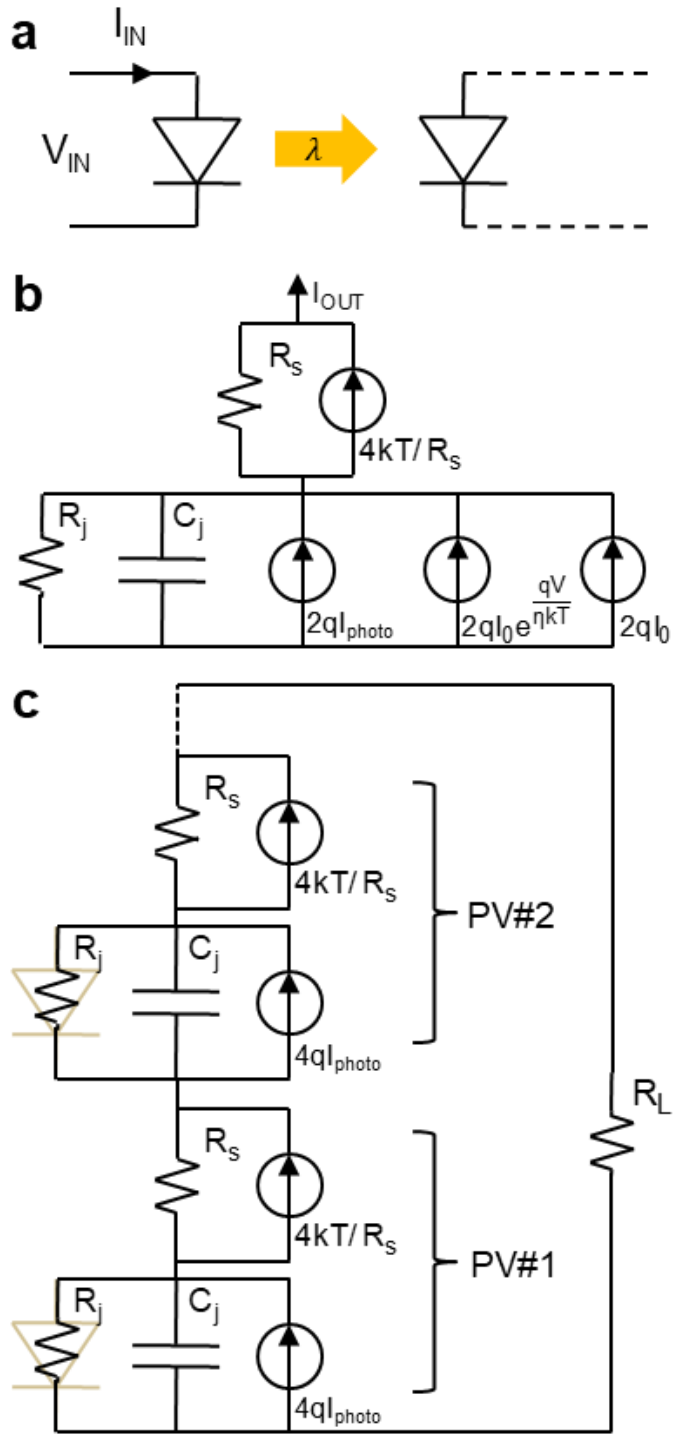

Extended Data Figure 9. **Noise analysis for the photonic transformer circuit. a**, A circuit consisting of one LED and one PV cell. **b**, The noise model for the circuit shown in a. **c**, Noise model for a photonic transformer circuit containing N LEDs and N PV cells.

|  | LED | PV |
|---|---|---|
| active region material | GaAs | Si |
| bandgap | 1.424 eV | 1.12 eV |
| SRH lifetime ($\tau$) | 22 ns | 92.8 ns |
| radiative recombination coefficient ($B$) | $1.3\times10^{-10}$ cm$^3$/s | $1\times10^{-14}$ cm$^3$/s |
| Auger coefficient ($C$) | $9.1\times10^{-30}$ cm$^6$/s | $1.3\times10^{-30}$ cm$^6$/s |
| active region area ($A$) | 0.2 mm × 0.2 mm | 2.65 mm × 2.65 mm |
| active region thickness ($t$) | 39.3 nm | 1.09 μm |
| series resistance ($R_s$) | 1.47 mΩ·cm$^2$ | 2.34 Ω·cm$^2$ |
| LED light extraction efficiency ($\eta_{EXT}$)/PV EQE ($\eta_{RES}$) | 0.41 | 0.897 |
| doping level | intrinsic | $N_d = 1.8\times10^{15}$ cm$^{-3}$ |
| view factor $f_{LED->PV}$ | 0.73 | |

Extended Data Table 1. **Parameters of the LED and the PV cell from measurement and fitting the experimental data.** The active region area and the sensitivity of the PV cell are provided by the product datasheets. The bandgaps are obtained from Ref. [52]. We note that the radiative recombination coefficient and the Auger coefficient for PV cells are directly taken from Ref. [52] since they have negligible effects on the performance of the device. The series resistances include resistances at both board level and package level, as well as resistances within the semiconductors.

|  | LED | PV |
|---|---|---|
| SRH lifetime ($\tau$) | 16.7 μs | 35 ms |
| radiative recombination coefficient (B) | $7\times10^{-10}\,\text{cm}^3/\text{s}$ | same as in Table 1 |
| Auger coefficient (C) | $3.5\times10^{-30}\,\text{cm}^6/\text{s}$ | same as in Table 1 |
| series resistance ($R_\text{s}$) | same as in Table 1 | $0.32\,\Omega\cdot\text{cm}^2$ |
| LED light extraction efficiency ($\eta_\text{EXT}$)/PV EQE ($\eta_\text{RES}$) | 0.8 | same as in Table 1 |
| view factor $f_\text{LED->PV}$ | 1 ||

Extended Data Table 2. **Parameters of the LED and the PV cell used in Extended Data Figure 7.** The PV parameters are obtained from Ref. [43]. The LED parameters are from Ref. [44], with the light extraction efficiency obtained from the product sheet of a high-efficiency infrared 850 nm LED (Osram LZ1-00R402).